\documentstyle[12pt]{article}
\newcommand{\cjt}{\widetilde{\cal J}}
\newcommand{\cft}{\widetilde{{\cal F}_t}}
\newcommand{\ccb}{\bar{\cal C}}

\newcommand{\cj}{{\cal J}}
\newcommand{\cf}{{\cal F}_t}
\newcommand{\cc}{{\cal C}}
\newcommand{\cee}{{\cal E}}
\newcommand{\jev}{{\cal J}_{ev}}
\newcommand{\jod}{{\cal J}_{od}}
\newcommand{\cg}{{\cal G}}

\newcommand{\sbb}{\bar{s}}
\newcommand{\bsb}{\bar{\bf s}}

\newcommand{\bdt}{{\bf d}_t}
\newcommand{\bs}{{\bf s}}
\newcommand{\bc}{{\cal C}}
\newcommand{\bcb}{\bar{{\cal C}}}

\newcommand{\bd}{{\bf d}}
\newcommand{\bdm}{{\bf d}_M}

\begin{document}
\baselineskip 22pt

\vspace*{0.8cm}

\begin{center}
{\large \bf BRST Quantization of Gauge Theory
      in Noncommutative Geometry: Matrix Derivative
       Approach}

\vspace{0.8cm}

Chang-Yeong Lee and
Dae Sung Hwang  \\
{\it Department of Physics, Sejong University, Seoul
    133-747, Korea} \\

\vspace*{0.2cm}

Yuval Ne'eman${}^{\#}$ \\
{\it Wolfson Distinguished Chair of Theoretical Physics,
         Raymond and Beverly Sackler Faculty of Exact Sciences,
        Tel Aviv University,
      Tel Aviv 69978, Israel}

\vspace{0.8cm}

{\bf Abstract} \\

\end{center}

The BRST quantization of a gauge theory in noncommutative geometry is
carried out in the ``matrix derivative" approach. BRST/anti-BRST
transformation rules are obtained by applying the horizontality
condition, in the superconnection formalism. A BRST/anti-BRST invariant
quantum action is then constructed, using an adaptation of the
method devised by Baulieu and Thierry-Mieg for the Yang-Mills case.
The resulting quantum action turns out to be the same as that of a
gauge theory in the 't Hooft gauge with spontaneously broken symmetry.
Our result shows that only the even part of the supergroup acts as a
gauge symmetry, while the odd part effectively provides a global symmetry.
We treat the general formalism first, then work out the $SU(2/1)$ and
$SU(2/2)$ cases explicitly.\\

\vfill

\noindent
PACS number(s): 02.40.-k, 12.10.-g, 12.15.Cc  \\
\hbox to 6cm{\hrulefill}  \\
\baselineskip 12pt
\noindent
\begin{itemize}
\item[${}^{\#}$]{Also on leave from: Center for Particle Physics,
  University of Texas, Austin, Tx 78712, USA}
\end{itemize}

\thispagestyle{empty}
\pagebreak

\baselineskip 22pt
\noindent
{\bf \large I. Introduction}\\
The Higgs mechanism makes it possible to give masses to gauge bosons,
while preserving the gauge symmetry. In this construction, some of the
original scalar particle fields `mutate' into the longitudinal components of
the (now massive) gauge bosons. This fact may reflect the existence of an
underlying structure, in which the gauge bosons and the original scalar
particles belong to the same multiplet of a larger group. It is, therefore,
natural to search for such a larger symmetry group and a suitable multiplet.
As a matter of fact, this idea was implemented many years ago, using the
supergroup $SU(2/1)$ \cite{yn}; it was also shown that this use of a
supergroup could be extended to a large class of spontaneously broken
symmetries \cite{nt}. More recently, the idea has further mathematically
evolved within the {\it superconnection} construct \cite{ntm,ntm2,ns,ln}.

Another recent advance in mathematical pysics has consisted \cite{conn} in
A. Connes' {\it noncommutative geometry}. In this formalism, the Dirac
K-cycle on a star algebra acting on a Hilbert space, plays an important role,
with possible applications to particle physics. Connes and Lott \cite{conlot}
then showed in particular that the {\it standard model} could be obtained in
noncommutative geometry, as a gauge theory with a built-in spontaneous
symmetry breakdown mechanism. Their work has been further extended to
GUT (grand unified theories) \cite{chamff1}, to gravity \cite{chamff2},
and to supersymmetric theories \cite{cham}.

Soon after the work of Connes and Lott, Coquereaux and other workers
\cite{couqet,sch} showed that the Connes-Lott approach is equivalent to
a theory based on the superconnection concept \cite{ns,qui},
rediscovering SU(2/1) in
the process. In Coquereaux {\it et al.}'s formulation,
a $Z_{2}$ graded space of matrix-valued forms is constructed, with a
generalized derivative; 0-form and 1-form fields together represent a
superconnection. The generalized derivative consists of the usual Cartan
exterior differential operator, raising the form degree by one unit and
thus also changing its Grassmann grading (which we denote as `w-grading',
i.e. {\bd} has odd w-grading) plus a graded {\it discrete} operator
consisting in a (graded) commutator with a constant matrix and satisfying
certain algebraic conditions (including odd grading in a supergroup's
generating superalgebra, `g-odd' in our nomenclature). This graded
commutator (or supercommutator) with a constant matrix is the
{\it matrix derivative} \cite{sch}.
We shall denote the Coquereaux {\it et al.}
approach as {\it the matrix derivative} approach.

The equivalence between the Connes-Lott and Coquereaux {\it et al.}
approaches
has been stressed by Scheck and collaborators \cite{pps}. In both
approaches, the 0-form scalar field is interpreted geometrically as an object
interconnecting a two-sheeted world, whereas the 1-form field plays the usual
role of a gauge field. The end-product is equivalent to an extension of the
internal supersymmetry method in its superconnection formulation, completing,
as we shall see, its geometric generation of a spontaneous symmetry breakdown
mode for a local gauge symmetry.

We have recently quantized the SU(2/1) electro-weak theory in the
superconnection formalism \cite{hln}. As an extension of this work, we
now include in the present paper the quantization of the noncommutative
geometry version of this "supergauge theory", by adjoining the matrix
derivative approach to the superconnection formulation. Actually, this
formulation goes beyond the internal supersymmetry method in one aspect,
namely the emergence of the {\it negative squared mass} term for
the scalar (Higgs) field from the geometry; in our previous treatment,
most terms in the spontaneous symmetry breakdown Lagrangian emerged
geometrically, namely (aside from the usual Yang-Mills term) the `free'
Higgs field Lagrangian plus its interaction with the gauge bosons -- and
the quartic Higgs field potential; the exception, which had to be put in
`by hand' (and thus also broke the symmetry explicitly) was this negative
squared mass term, which is now provided by the matrix derivative.

We obtain the BRST/anti-BRST transformation  rules of the theory,
applying our {\it horizontality} condition, extending Thierry-Mieg's ansatz
\cite{ln,tm,tmn}. We construct the quantum action by adapting the
Baulieu/Thierry-Mieg method \cite{btm} for
the Yang-Mills theory.

There are two important features deriving from our result. The first is the
fact that we obtain the most appropriate gauge condition for a
spontaneously broken gauge theory with scalar field, the 't Hooft gauge
\cite{tht,fls}, simply by adapting the method of Ref. \cite{btm},
which would give the
 Landau gauge for the unbroken
Yang-Mills theory,
 to the noncommutative geometry framework.
The other relates to the physical content of a gauge theory in the
noncommutative setting. Our quantization reveals that only the even part
of the supergroup indeed acts as a gauge symmetry; the odd part simply
produces a global symmetry. The resulting BRST transformation rules
for the fields are thus the same as those of the spontaneously broken gauge
theory with a Higgs mechanism, except that the scalar field transformation
rule is changed by the addition of a constant shift (a vacuum shift), due
to the action of the matrix derivative, thereby implementing geometrically
the triggering of the spontaneous breakdown. Other fields are not affected
by the appearance of the matrix derivative.

In section 2, we study the BRST quantization in the matrix derivative
approach for the general case. In section 3, we treat the $SU(2/1)$ gauge
theory, effectively an algebraically constrained standard model
$SU(2)\times U(1)$ gauge theory of the electro-weak interaction.
In section 4, we consider an $SU(2/2)$ gauge theory, which reduces to the
spontaneously broken symmetry of an $SU(2)\times SU(2)$ $\sigma$-model.
Section 5 contains a discussion and conclusions. \\

\noindent
{\bf \large II. BRST/anti-BRST symmetry and quantum action}\\

In the matrix derivative approach of a noncommutative geometrical gauge
theory, the 0-form scalar field and 1-form gauge field together form a
superconnection, with w-odd forms in the g-even part and w-even forms in
the g-odd part of the supergroup.
We write the superconnection $\cj\ $  as
%eq. 1
\begin{equation}
\cj\ = \jev\ + \jod\ = \left(\matrix{\omega_0&0\cr 0&\omega_1\cr} \right)
+\left(\matrix{0&L_{01}\cr L_{10}&0\cr}\right).
\label{c1}
\end{equation}
The overall $Z_{2}$ grading is given by the sum of the supermatrix grading
($Z_{2}$ `g'-grading) and the differential form grading ($Z_{2}$
`w'-grading). The total grading of the superconnection is therefore odd,
in this $Z_{2}$ graded space \cite{hln}. Multiplication in this superspace
is given by \cite{ns,couqet}
%eq. 2
\begin{equation}
(h\otimes W)\cdot (h^\prime \otimes W^\prime )
=(-1)^{\vert W\vert  \vert h^\prime\vert }
(h h^\prime )\otimes (W W^\prime ),
\label{c2}
\end{equation}
where $W,\ W'$ are differential forms of fixed Grassmannian $Z_{2}$
w-gradings $\vert W \vert , \ \vert W' \vert $, and $h,\ h' $ are
supermatrices of fixed $Z_{2}$ g-grading $\mid h \mid ,\  \mid h' \mid $.
With this convention, we obtain the product rule for any two elements
in our total $Z_{2}$ graded space, assuming $A,B,C,D $ to be matrix-valued
differential forms, which have fixed $Z_{2}$ w-gradings of 0 or 1,
depending on whether they are even or odd forms, respectively,
\cite{ns,couqet}
%eq 3
\begin{equation}
\left( \begin{array}{cc}A & B\\ C & D \end{array} \right) \cdot
\left( \begin{array}{cc}A' & B' \\ C' & D' \end{array} \right) =
\left( \begin{array}{cc} A \wedge A' + (-1)^{ \mid {\rm B} \mid }
 B \wedge  C' & (-1)^{ \mid {\rm A} \mid } A \wedge B' +
 B \wedge D' \\
  C \wedge A' +  (-1)^{ \mid {\rm D} \mid }
 D \wedge C' & (-1)^{ \mid {\rm C} \mid }
 C \wedge B' + D \wedge D' \end{array} \right).
\label{c3}
\end{equation}

Once the superconnection is given, the supercurvature $\; \cf\ $ is defined
in the usual manner, with the generalized derivative
$\; \bdt\ $, consisting of the usual 1-form differential operator
$\; \bd\  $ and the {\it matrix derivative} $\; \bdm\ $ \cite{couqet,sch}:
%eqs.4,5&6
\begin{eqnarray}
\cf\ & = & \bdt\  \cj\ + \cj\ \cdot \cj\ , \label{c4} \\
\bdt\ & = & \bd\ + \bdm\ , \label{c5} \\
\bd\ & = &  \left(\matrix{d&0\cr 0&d\cr}\right), \; \;
{\rm where} \; \;
 d = 1\otimes dx^{\mu} \frac{\partial}{\partial x^{\mu}}.
\label{c6}
\end{eqnarray}
The matrix derivative is given by
%eq.7
\begin{equation}
 \bdm\ = i [ \eta , \ \ \ ]_{\pm}, \; \; {\rm where} \; \; \eta =
 \left(\matrix{0& \zeta \cr \overline{\zeta} &0\cr}\right).
\label{c7}
\end{equation}
Here $ \zeta $ and $\overline{\zeta}$ are constant matrices of zero forms,
satisfying
%eq.8
\begin{equation}
\overline{\zeta} \zeta=\zeta \overline{\zeta} \propto 1
, \label{c8}
\end{equation}
so that the matrix derivative satisfies the nilpotency condition,
$ {\bdm\ }^{2}= 0$. Note that the total grading of the matrix
derivative $\; \bdm\ $is odd. Thus the matrix derivative is a
supercommutator, i.e. it acts as a commutator for objects of even
total grading and as an anticommutator for objects of odd total
grading, where by `total', we mean the product of the
gradings of 'g' and 'w'. \\

We now write the classical action of the gauge theory in noncommutative
geometry as
%eq.9
\begin{equation}
{\cal S}_{cl} = - \frac{1}{4} \int {\rm Tr} \; {\cf\ }^{\star} \cdot \cf\ ,
\label{c9}
\end{equation}
where $\star $ denotes taking the Hermitian conjugate for
supermatrices and taking the Hodge dual for differential forms.
In order to find the BRST/anti-BRST transformation rules,
we use the so-called {\it horizontality} condition\cite{ntm,tm,tmn,btm},
which is another description of the Maurer-Cartan equation:
%eq.10
\begin{equation}
 \cft\ = \cf\ ,
\label{c10}
\end{equation}
 where  $\cft\ $ is the supercurvature, defined in the extended space of
the {\it doubled} fiber bundle \cite{hln},
%eq.11
\begin{equation}
\cft\  = {\widetilde{{\bf d}_t}}\ \cjt\ + \cjt\ \cdot \cjt\ .
\label{c11}
\end{equation}
`Doubling' implies the extension of the base manifold through doubling the
fiber, from  $\{\cg\}$ to $\{\cg\}\otimes \{\cg\}$,
so that we have a gauge fiber
coordinate $y$ and its dual $\bar{y} \ $ \cite{ln,tm,tmn,btm}.
In this extended
space, the generalized derivative and superconnection are given by
%eq.12
\begin{eqnarray}
 {\widetilde{{\bf d}_t}} & = &  {\bdt\ } + {\bf s} + \bsb\ ,
 \label{c12} \\
\cjt\ & = & {\cal J} + {\cal C} + \ccb\
\label{c12a}.
\end{eqnarray}
Here,
${\bf s}$ and $\bsb\ $ are 1-form differential operators acting respectively
on the coordinates of the fiber and of its dual:
%eq. 13
\begin{eqnarray}
{\bf s} & = &  \left(\matrix{s&0\cr 0&s\cr}\right) \; \;
{\rm where} \; \;
 s = 1 \otimes dy^{N} \frac{\partial}{\partial y^{N} } , \nonumber \\
 \bsb\ & = & \left( \begin{array}{cc} \sbb\  & 0 \\
    0 & \sbb\ \end{array} \right) \; \;
{\rm where} \; \;
 \sbb\ = 1 \otimes d {{\bar{y}} }^{M} \frac{\partial}
  {\partial {{\bar{y}}}^{M}}.
\label{c13}
\end{eqnarray}
 $ {\cal C}$ and $\ccb\ $ are obtained from ${\cal J}$ by replacing
$dx^{\mu} $ by $ dy^{N}$ and $d {{\bar{y}} }^{M}$, and
represent the ghost and anti-ghost fields, respectively:
%eq.14
\begin{eqnarray}
\cc\ & = & \left( \begin{array}{cc} c_{0N}dy^{N} & 0 \\
         0 & c_{1N}dy^{N} \end{array} \right) \equiv \left(
         \begin{array}{cc} c_0 & 0 \\ 0 & c_1 \end{array} \right)
       ,  \nonumber \\
\ccb\ & = & \left( \begin{array}{cc} {\bar{c_0}}_{M}d {{\bar{y}} }^{M} & 0 \\
         0 &  {\bar{c_1}}_{M} d {{\bar{y}} }^{M} \end{array} \right)
         \equiv \left(
        \begin{array}{cc} {\bar{c_0}}  & 0 \\ 0 & {\bar{c_1}} \end{array}
        \right).
        \label{c14}
\end{eqnarray}

After applying the horizontality condition we  obtain the BRST/anti-BRST
transformation rules:
%eq.15
\begin{eqnarray}
(dy)^{1} & : &
\bs\  \cj\  =  -\bdt\  \bc\
- \cj\ \cdot \bc\ -\bc\ \cdot \cj\ , \nonumber \\
(d{\bar{y}})^{1} & : &
\bsb\  \cj\  =  -\bdt\  \bcb\
- \cj\ \cdot \bcb\ -\bcb\ \cdot \cj\ ,  \nonumber \\
(dy)^{2} & : &
\bs\  \bc\  =  -\bc\ \cdot \bc\ ,  \label{c15} \\
(d{\bar{y}})^{2} & : &  \bsb\  \bcb\  =  -\bcb\ \cdot \bcb\ , \nonumber \\
(dy)^1(d{\bar{y}})^{1} & : &
\bs\  \bcb\ + \bsb\  \bc\ + \bc\ \cdot \bcb\
+ \bcb\ \cdot \bc\  =  0 .
\nonumber
\end{eqnarray}
By introducing an auxiliary field $ \cee\ $ such that
%eq.16
\begin{equation}
\bs\ \ccb\ \equiv \cee\ , \ \ {\rm i.e.,} \ \
  \left( \begin{array}{cc} s {\bar{c_0}} & 0 \\ 0 & s {\bar{c_1}} \end{array}
  \right) \equiv \left( \begin{array}{cc} b_0 & 0 \\ 0 & b_1 \end{array}
  \right),
\label{c16}
\end{equation}
we can fix the remaining BRST/anti-BRST transformation rules,
%eq.17
\begin{eqnarray}
\bsb\  \bc\ & = & -\cee\ -\bc\ \cdot \bcb\ - \bcb\ \cdot \bc\ , \nonumber \\
\bs\  \cee\ & = & 0 ,
\label{c17} \\
\bsb\  \cee\ & = & -\bsb\  (\bc\ \cdot \bcb\ +\bcb\ \cdot \bc\ )
=-\bcb\ \cdot \cee\ + \cee\ \cdot \bcb\ .
\nonumber
\end{eqnarray}
One can easily check the nilpotency property of the BRST/anti-BRST
transformations,  $ {\bs\ }^{2} = {\bsb\ }^{2} = 0 $,
for the above transformation rules (\ref{c15}), (\ref{c16}) and
(\ref{c17}).

Decomposing $\cj\ $ into ${\cj\ }_{ev} + {\cj\ }_{od} $
as in (\ref{c1}), we can write the even and odd parts of the first
two equations in (\ref{c15}) separately as follows, by noting that
$\bd\ , \  \ \bs\ $ and $\bsb\ $ are even matrices, whose entries are
one-form differential operators.
%eq.18
\begin{eqnarray}
{\rm even \ \ part} & : &
\bs\  \jev\  =  -\bd\  \bc\
- \jev\ \cdot \bc\ -\bc\ \cdot \jev\ , \nonumber \\
      &  &
\bsb\  \jev\  =  -\bd\  \bcb\
- \jev\ \cdot \bcb\ -\bcb\ \cdot \jev\ , \label{c18} \\
{\rm odd \ \ part} & : &
\bs\  \jod\  =  -\bdm\  \bc\
- \jod\ \cdot \bc\ -\bc\ \cdot \jod\  , \nonumber \\
      &   &
\bsb\  \jod\  =  -\bdm\  \bcb\
- \jod\ \cdot \bcb\ -\bcb\ \cdot \jod\ .
\nonumber
\end{eqnarray}
Note that the even parts are the usual BRST/anti-BRST
transformation rules of a one-form gauge field \cite{btm},
while the odd parts are those of a matter field, plus the additional
terms caused by the matrix derivative. These additional terms
represent a translation of the scalar field and correspond to the
vacuum shift in the usual Higgs mechanism. The difference, however, is
that this is a built-in property of a gauge theory in the noncommutative
geometry setting, in contradistinction to the conventional Higgs construction.
The system's `ordinary' gauge symmetry is thereby broken explicitly through
that geometrical setting.\\

Adapting the Baulieu/Thierry-Mieg method for
a BRST/anti-BRST invariant quantum action, which yields the Landau
gauge for the usual Yang-Mills theory\cite{btm}, we write the quantum
action as
%eq{c19}
\begin{equation}
{\cal S}_{Q} = - \frac{1}{4} \int Tr \; \{  {\cf\ }^{\star} \cdot \cf\
       - \bs\  \bsb\  ({\cj\ }^{\star} \cdot \cj\ )
       + \alpha \ \ \bs\  ( {\bcb }^{\star} \cdot \cee\ ) \},
\label{c19}
\end{equation}
where, $\alpha $ is a parameter.
Using the transformation rules (\ref{c15}), (\ref{c16}), (\ref{c17})
and (\ref{c18}), we obtain
%eq.20
\begin{equation}
Tr \{ \bs\  \bsb\  ( {\jev\ }^{\star}\cdot \jev\ ) \}
         =  2 \ Tr\{ ( \jev\ )^{\star} \cdot ( \bd\  \cee\ ) +
         ( {\bd\ \bcb\ } )^{\star} \cdot
         (\bd\  \bc\ + \jev\ \cdot \bc\ + \bc\ \cdot \jev\ )
         \} ,
\label{c20}\\
\end{equation}
%eq.21
\begin{equation}
Tr \{ \bs\  \bsb\  ( {\jod\ }^{\star}\cdot \jod\ ) \}
         =  2 \ Tr \{ ( \jod\ )^{\star} \cdot ( \bdm\  \cee\ ) +
         ( \bdm\  \bcb\ )^{\star} \cdot
         (\bdm\  \bc\ + \jod\ \cdot \bc\ + \bc\ \cdot \jod\ ) \} ,
\label{c21}
\end{equation}
and
%eq.22
\begin{equation}
Tr \left\{ \alpha \bs\  ( {\ccb\ }^{\star}\cdot \cee\ ) \right\}
    = Tr \left\{ \alpha {\cee\ }^{\star} \cdot \cee\ \right\} .
\label{c22}
\end{equation}
Thus, the quantum action $\; {\cal S}_{Q} \; $ can be written as
%eq.23
\begin{eqnarray}
{\cal S}_{Q} &=& - \frac{1}{4}
       \int  Tr \; \{ {\cf\ }^{\star}\cdot \cf\
       + \alpha \ \  {\cee\ }^{\star} \cdot \cee\
\nonumber\\
       &-& 2 ( \jev\ )^{\star} \cdot ( \bd\  \cee\ ) - 2
         ( {\bd\ \bcb\ } )^{\star} \cdot
         (\bd\  \bc\ + \jev\ \cdot \bc\ + \bc\ \cdot \jev\ )
\label{c23}\\
       &-& 2 ( \jod\ )^{\star} \cdot ( \bdm\   \cee\ ) - 2
         ( \bdm\  \bcb\ )^{\star} \cdot
         (\bdm\  \bc\ + \jod\ \cdot \bc\ + \bc\ \cdot \jod\ ) \} .
\nonumber
\end{eqnarray}
One can check that this quantum action is  BRST/anti-BRST invariant.

In the above quantum action (\ref{c23}), the terms with  the auxiliary
field $\cee\ $ are the gauge fixing terms and give rise to the
't Hooft gauge condition \cite{tht,fls} as we shall see in the next
two sections. The first term is the classical action, and the remaining terms
constitute the kinetic and interaction terms of the ghost fields.
In the following two sections we calculate the quantum action (\ref{c23})
for the  $SU(2/1)$ and $SU(2/2)$ cases explicitly. \\

\noindent
{\bf \large III. BRST quantization of the $SU(2/1)$ case}\\

The generators of $SU(2/1)$  are the same as those of $SU(3)$,
namely the conventional $\lambda$ - matrices, except for $t_8$,
which is given by
%eq.24
\begin{equation}
t_{8}= \frac{1}{\sqrt{3}} \left( \begin{array}{ccc}
   -1 & 0  & 0 \\ 0 & -1 & 0 \\ 0 & 0 & -2 \end{array} \right) ,
\label{c24}
\end{equation}
in order to  satisfy $STr(t_i)=0$.
We write the $SU(2/1)$ superconnection as
%eq.25
\begin{eqnarray}
\cj\ &=&i{t_i} {J_i} \; \ (i=1,2,\cdots ,8)  \nonumber\\
 &=&\jev\ + \jod\
=i\left( \matrix{{\tau_a}{W_a} -\frac{1}{\sqrt{3}} B&0\cr
0&-\frac{2}{\sqrt{3}} B\cr}\right)
+i\left( \matrix{0&{\sqrt{2}}\Phi \cr {\sqrt{2}}{\Phi}^{\dag} &0\cr}\right),
\label{c25}
\end{eqnarray}
where we identified the gauge and Higgs fields
$W_a ,B,\Phi ,$ and ${\Phi}^{\dag}$ with the components;
$\ W_a =J_a \ (a=1,2,3), \; B=J_8 , \; \Phi =\frac{1}{\sqrt{2}}\left(
\begin{array}{c} J_4 -iJ_5 \\ J_6 -iJ_7 \end{array}
\right) , \ $ and
${\Phi}^{\dag} =\frac{1}{\sqrt{2}}\left(
\begin{array}{c} J_4  +iJ_5 \\ J_6 +iJ_7 \end{array}
\right) .$

We now introduce the ghost, anti-ghost, and auxiliary fields, in the
doubled-fiber bundle space.
%eq.26
\begin{eqnarray}
\cc\ &=&i\left( {\matrix{\tau_a c_a -\frac{1}{\sqrt{3}} c_8 &0\cr
0&-\frac{2}{\sqrt{3}} c_8\cr}}\right),
\; \ \ccb\ =i\left( {\matrix{\tau_a {\bar{c_a}}
-\frac{1}{\sqrt{3}} {\bar{c_8}} &0\cr
0&-\frac{2}{\sqrt{3}} {\bar{c_8}}\cr}}\right),\nonumber\\
\cee\ &=&i\left( {\matrix{{\tau_a} {b_a} -{\frac{1}{\sqrt{3}}} b_8 &0\cr
0&-{\frac{2}{\sqrt{3}}} b_8\cr}}\right)  \ \ \  ( a=1,2,3 ).
\label{c26}
\end{eqnarray}
In order to derive the  BRST/anti-BRST transformation rules, we apply
eqs.(\ref{c15})-(\ref{c18}) of the previous section. In calculating the
$SU(2/1)$ case, we encounter the following difficulty. With the
$3 \times 3$ matrix representation, it is not possible to choose a
constant matrix
$\eta = \left(\matrix{0& \zeta \cr \overline{\zeta} &0\cr}\right)$
for the matrix derivative, satisfying the condition (\ref{c8}),
$ \; \overline{\zeta} \zeta=\zeta \overline{\zeta} \propto 1$,
which is essential for the nilpotency of the matrix derivative. In order
to resolve this difficulty,  we first extend all $3 \times 3$ matrix
representations of fields into $4 \times 4$ matrices, simply by adjoining
a 4th row and a 4th column, with all components vanishing.
We then choose the $\eta $ matrix in this extended $4 \times 4$ matrix
representation space, in which it does satisfy the nilpotency condition.
This $4 \times 4$ $\eta$ matrix, enables us to perform all calculations
involving the  $\eta$ matrix, such as evaluating the supercurvature, etc.
After this is done, we project back onto the $3 \times 3$ matrix
representation space, simply discarding the 4th row and column. Note that
this construction reflects the fact that the true fundamental
representation of SU(2/1) is 4-dimensional \cite{ntm}, reflecting the
homomorphism with OSp(2/2) and fitting the internal quantum numbers for
quarks, i.e. $(u_{R}/u_{L},d_{L}/d_{R})$ where the order follows descending
weak hypercharges $(4/1,1/-2)$ (in units of $(1/3)$). However, for integer
charges, the upper state trivializes and disconnects (e.g. the $\nu_{R}$)
and we are left with the 3-dimensional representation. As a matter of fact,
the procedure we use here also corresponds to the {\it projective module}
method of Connes and Lott \cite{conlot}.
We thus perform the actual calculation with
\[ \zeta =\overline{\zeta}= \sqrt{2} k
 \left(\matrix{0& 1 \cr 1 &0\cr}\right), \; \; \; k:{\rm real}, \]
 and obtain
the following BRST/anti-BRST transformation rules.
%eq.27
\begin{eqnarray}
s A_{II} & = & -d c_{II} - A_{II} c_{II} -c_{II} A_{II}
\nonumber , \\
\bar{s} A_{II} & = & -d \bar{c_{II}} - A_{II}\bar{c_{II}} -\bar{c_{II}} A_{II}
 , \nonumber \\
s A_I &=& -d c_I ,  \ \  \bar{s} A_I = -d \bar{c_I}
 , \nonumber\\
s \Phi & = & -c_{II} (\Phi + \xi ) -\frac{1}{\sqrt{3}} c_I (\Phi
+ \xi )
\nonumber , \\
\bar{s} \Phi &=& -\bar{c_{II}} (\Phi + \xi )
-\frac{1}{\sqrt{3}} \bar{c_I}( \Phi
+ \xi )  \nonumber , \\
s \Phi^{\dag} & = & (\Phi^{\dag} +\xi^{\dag} ) c_{II}
+ \frac{1}{\sqrt{3}} (\Phi^{\dag} + \xi^{\dag} ) c_I
\nonumber , \\
\bar{s} \Phi^{\dag} &=& (\Phi^{\dag} +\xi^{\dag} ) \bar{c_{II}}
+ \frac{1}{\sqrt{3}} (\Phi^{\dag}
+ \xi^{\dag} ) \bar{c_I}
\nonumber , \\
s c_{II} & = & -c_{II} c_{II} ,  \ \
\bar{s} \bar{c_{II}} = -\bar{c_{II}} \bar{c_{II}} ,
\label{c27}\\
s c_I &=& \bar{s} \bar{c_I} = 0
\nonumber , \\
s {\bar{c_{II}}} & = & b_{II} , \ \
\bar{s} c_{II} = -b_{II} -c_{II} \bar{c_{II}} -\bar{c_{II}} c_{II}
\nonumber , \\
s b_{II} & = & 0  ,  \ \
\bar{s} b_{II} = -\bar{c_{II}} b_{II} + b_{II}\bar{c_{II}} ,
\nonumber  \\
s \bar{c_I} &=& - \bar{s} c_I = b_I , \ \
s b_I = \bar{s} b_I = 0
\nonumber ,
\end{eqnarray}
where
%eq.28
\begin{eqnarray}
A_{II} &=& i {\tau}_a W_a , \ \ A_I = iB , \ \ c_{II} =i {\tau}_a c_a ,
\ \ c_I =i c_8 ,
\nonumber\\
\bar{c_{II}} &=& i {\tau}_a \bar{c_a} , \ \  \bar{c_I} = i\bar{c_8} , \ \
b_{II} = i {\tau}_a b_a \ \ (a=1,2,3) , \ \ b_I = i b_8
  \label{c28} , \\
\xi & = & k \left( \begin{array}{c} 0 \\ 1 \end{array} \right).
\nonumber
\end{eqnarray}
Note that the transformation rules of $\Phi$ and ${\Phi}^{\dag}$
correspond to those of the Higgs fields with a shifted vacuum.
For the supercurvature we obtain
%eq.29
\begin{equation}
\cf\ =\left(
\matrix{F_W -{\frac{1}{\sqrt{3}}}F_B -2(\Phi {\Phi}^{\dag}
+\xi {\Phi}^{\dag} +\Phi {\xi}^{\dag} )& \ \ -i{\sqrt{2}}
(D \Phi +(i{\vec{W}}\cdot {\vec{\tau}} +{\frac{i}{\sqrt{3}}} B)\xi )
\cr -i{\sqrt{2}}
(D {\Phi}^{\dag} - {\xi}^{\dag} (i{\vec{W}}\cdot {\vec{\tau}}
+{\frac{i}{\sqrt{3}}} B) ) & \ \ -{\frac{2}{\sqrt{3}}}F_B
-2({\Phi}^{\dag} \Phi +{\xi}^{\dag} \Phi +{\Phi}^{\dag} \xi )\cr}
\right) ,
\label{c29}
\end{equation}
where
%eq.30
\begin{eqnarray}
& & F_W = {\frac{1}{2}} F_{W\mu\nu} dx^{\mu}\wedge dx^{\nu}
= d(i{\vec{W}}\cdot {\vec{\tau}} )
+ (i{\vec{W}}\cdot {\vec{\tau}} )(i{\vec{W}}\cdot {\vec{\tau}} ) ,
\nonumber\\
& & F_B = {\frac{1}{2}} F_{B\mu\nu} dx^{\mu}\wedge dx^{\nu} = d(iB) ,
\label{c30}\\
& & \Phi =
\left( \begin{array}{c} {\phi}^+ \\ {\phi}^0 \end{array} \right)
={\frac{1}{\sqrt{2}}}
\left( \begin{array}{c} {\phi}_3 + i{\phi}_4 \\ {\phi}_1 + i{\phi}_2
\end{array} \right)
={\frac{1}{\sqrt{2}}}
\left( \begin{array}{c} J_4 - iJ_5 \\ J_6 - J_7
\end{array} \right) ,
\nonumber\\
& &D \Phi
=(D \Phi )_{\mu} dx^{\mu} =
d\Phi + (i{\vec{W}}\cdot {\vec{\tau}} +
{\frac{i}{\sqrt{3}}} B ) \Phi , \nonumber\\
& &D {\Phi}^{\dag}
=(D {\Phi}^{\dag})_{\mu} dx^{\mu} =
d{\Phi}^{\dag} -
{\Phi}^{\dag} (i{\vec{W}}\cdot {\vec{\tau}}
+ {\frac{i}{\sqrt{3}}} B ) .
\nonumber
\end{eqnarray}
We use $d^4 x =dx^0\wedge dx^1\wedge dx^2\wedge dx^3,\ \
\epsilon_{0123} = 1,$ and adopt the convention of Ref.\cite{egh}
for the dual of a differential form in $n$ dimension, required for
(\ref{c23}),
%eq.31
\begin{equation}
\ast (dx^{i_1}\wedge dx^{i_2}\wedge\cdots\wedge dx^{i_p})
={\frac{1}{(n-p)!}}{\epsilon}^{i_1i_2\cdots i_p}{}_{i_{p+1}\cdots i_n}
dx^{i_{p+1}}\wedge\cdots\wedge dx^{i_n}
\label{c31}
\end{equation}
satisfying $\ast\ast {\omega}_p = (-1)^{p(n-p)} {\omega}_p$
for a $p$-form ${\omega}_p$.

Selecting the metric ${g_{\mu\nu}} = (-1,+1,+1,+1)$,
the first term in (\ref{c23}), the classical action, is given by
%eq.32
\begin{eqnarray}
{\cal L}_C &=& {\frac{1}{4}}{F_{W a \mu\nu}}{F_{W a}^{\mu\nu}}
+ {\frac{1}{4}}{F_{B\mu\nu}}{F_B^{\mu\nu}}\nonumber\\
& & - ( D {\Phi}^{\dag}
-{\xi}^{\dag} ( i{\vec{W}}\cdot {\vec{\tau}} +{\frac{i}{\sqrt{3}}} B ))_{\mu}
(D \Phi +
( i{\vec{W}}\cdot {\vec{\tau}} +{\frac{i}{\sqrt{3}}} B )\xi
 )^{\mu}\nonumber\\
& & - 2 (({\Phi}^{\dag} +{\xi}^{\dag} )(\Phi +\xi )-{\xi}^{\dag} \xi )^2
\nonumber\\
&=& {\frac{1}{4}}{F_{W a \mu\nu}}{F_{W a}^{\mu\nu}}
+ {\frac{1}{4}}{F_{B\mu\nu}}{F_B^{\mu\nu}}\nonumber\\
& & - ( D {\Phi}^{\dag} -ik({\sqrt{2}}W_- +{\frac{2}{\sqrt{3}}} Z))_{\mu}
(D \Phi + ik
\left(
\begin{array}{c} {\sqrt{2}} W_+ \\ {\frac{2}{\sqrt{3}}} Z \end{array}
\right) )^{\mu}\label{c32}\\
& & - 2 ({\Phi}^{\dag}\Phi + k ({\phi}^0 +{\bar{{\phi}^0}} ))^2 ,
\nonumber
\end{eqnarray}
where
$  W_{\pm}^{\mu} ={\frac{1}{\sqrt{2}}}({W_1^{\mu} \mp iW_2^{\mu}}) ,
Z^{\mu} = -{\frac{\sqrt{3}}{2}} W_3^{\mu} + {\frac{1}{2}} B^{\mu} ,
A^{\mu} = {\frac{1}{2}} W_3^{\mu} + {\frac{\sqrt{3}}{2}} B^{\mu} $.

In order to see the physical spectrum of the theory, we
now write the above expression in the {\it unitary gauge}, which is given by
 ${\Phi} = \left(
\begin{array}{c} 0 \\ {\frac{1}{\sqrt{2}}} \chi \end{array}
\right)$ with real $\chi$.
%eq.33
\begin{eqnarray}
{\cal L}_C^{UG} &=& {\frac{1}{4}}  {F_{W a \mu\nu}} {F_{W a}^{\mu\nu}}
+{\frac{1}{4}}{F_{B\mu\nu}} {F_B^{\mu\nu}}\nonumber\\
&-&\{ {\frac{1}{2}}{\partial}_{\mu}\chi {\partial}^{\mu}\chi
+g^2 W_{+\mu} W_-^{\mu} (\chi +{\frac{{\sqrt{2}}k}{g}} )^2
+ {\frac{2}{3}} g^2 Z_{\mu} Z^{\mu} (\chi +{\frac{{\sqrt{2}}k}{g}} )^2 \}
\label{c33}\\
&-& {\frac{1}{2}} ( g {\chi}^2 + 2{\sqrt{2}} k \chi )^2
\nonumber
\end{eqnarray}
The coupling constant $g$ is introduced by scaling the superconnection
as $\cj\ \rightarrow g \cj\ $. In this unitary gauge we see that only
one scalar field remains as a physical (and massive) Higgs field $\chi$,
whereas the other three scalars have been `mutated', now providing
the longitudinal components of $W_{\pm}$ and $Z$. The masses of the
massive particles are $M_{\chi} = 2 {\sqrt{2}} k,
M_{W} = {\sqrt{2}} k, M_Z = {\frac{2{\sqrt{2}}}{{\sqrt{3}}}} k,$
and we see  the relations ${\frac{M_{W}^2}{M_Z^2}} = {\frac{3}{4}}
= cos^2 {\theta}_W , \;  M_{\chi} = 2 M_{W} .$ We shall return to the
latter ratio ${\frac{M_{\chi}}{M_W}}$ in section 5, when discussing possible
quantum corrections.

We now write the quantum Lagrangian of (\ref{c23}) as
%eq.34
\begin{equation}
{\cal L}_Q = {\cal L}_C + {\cal L}_1 + {\cal L}_2 ,
\label{c34}
\end{equation}
where ${\cal L}_C$ is the classical Lagragian, ${\cal L}_1$ stands for
the ghost terms, and ${\cal L}_2$ for the gauge fixing terms.
After some calculations, we obtain ${\cal L}_1$,
%eq.35
\begin{eqnarray}
{\cal L}_1 &=& {\frac{1}{2}} \ \
tr [ {\partial}_{\mu} \bar{c_{II}} D^{\mu} c_{II}
    + 2 {\partial}_{\mu} \bar{c_{I}} {\partial}^{\mu} c_{I}
\nonumber\\
&+& \{
(\bar{c_{II}} + {\frac{1}{\sqrt{3}}}\bar{c_I} ) \xi ( {\Phi}^{\dag} +
{\xi}^{\dag} ) ({c_{II}} + {\frac{1}{\sqrt{3}}}{c_I} )\}
\label{c35}\\
&+& \{ {\xi}^{\dag} (\bar{c_{II}} + {\frac{1}{\sqrt{3}}}\bar{c_I} )
({c_{II}} + {\frac{1}{\sqrt{3}}}{c_I} ) (\Phi +\xi )
\} ] ,
\nonumber
\end{eqnarray}
where $D^{\mu} c_{II} = {\partial}^{\mu} c_{II} + [ A_{II}^{\mu} , c_{II} ] .$

For ${\cal L}_2$, we obtain
%eq.36
\begin{eqnarray}
{\cal L}_2
&=& {\frac{\alpha}{2}} \ \ \{ [(b_1)^2 +(b_2)^2 +(b_Z)^2 +(b_A)^2 ]
+ \frac{2}{\alpha} \ \
 [ b_1 ({\partial}_{\mu} W_1^{\mu} -{\sqrt{2}}k{\phi}_4 )
 \nonumber\\
&+& b_2 ({\partial}_{\mu} W_2^{\mu} -{\sqrt{2}}k{\phi}_3 )
+ b_Z ({\partial}_{\mu} Z^{\mu} - {\frac{2{\sqrt{2}}}{{\sqrt{3}}}} k
{\phi}_2 )
+ b_A ({\partial}_{\mu} A^{\mu} )] \} ,
\label{c36}
\end{eqnarray}
where $b_Z = -{\frac{\sqrt{3}}{2}} b_3 + {\frac{1}{2}} b_8 ,
b_A = {\frac{1}{2}} b_3 + {\frac{\sqrt{3}}{2}} b_8 .$
After integrating out the auxiliary fields $b_1 , b_2 , b_Z,$ and
$b_A$, $ \;  {\cal L}_2$ becomes
%eq.36a
\begin{equation}
{\cal L}_2 =-\frac{1}{2\alpha} \ \  \{
({\partial}_{\mu} W_1^{\mu} -{\sqrt{2}}k{\phi}_4 )^2
+({\partial}_{\mu} W_2^{\mu} -{\sqrt{2}}k{\phi}_3 )^2
+({\partial}_{\mu} Z^{\mu} - {\frac{2{\sqrt{2}}}{{\sqrt{3}}}} k
{\phi}_2 )^2
+ ( {\partial}_{\mu} A^{\mu} )^2 \} .
\label{c36a}
\end{equation}
This expression clearly shows that we obtain the gauge-fixed
quantum Lagrangian of the 't Hooft gauge \cite{tht,fls}, as we claimed in the
previous section:
%eq.37
\begin{eqnarray}
{\partial}_{\mu} W_1^{\mu} - M_W {\phi}_4 &=& 0 , \nonumber\\
{\partial}_{\mu} W_2^{\mu} - M_W {\phi}_3 &=& 0 , \nonumber\\
{\partial}_{\mu} Z^{\mu} -  M_Z
{\phi}_2 &=& 0 , \label{c37}\\
{\partial}_{\mu} A^{\mu} &=& 0 .  \nonumber
\end{eqnarray}

\noindent
{\bf \large IV. BRST quantization of $SU(2/2)$ case}\\

We now calculate the $SU(2/2)$ case.
The generators of $SU(2/2)$  are the same as those of $SU(4)$, except
for $t_8$ and $t_{15}$, which are replaced by
%eq.38
\begin{equation}
t_{8}= \frac{1}{\sqrt{3}} \left( \begin{array}{cccc}
   -1 & 0  & 0 & 0 \\ 0 & -1 & 0 & 0 \\ 0 & 0 & -2 & 0 \\ 0 & 0 & 0 & 0
   \end{array} \right) , \ \
t_{15}= \frac{1}{\sqrt{6}} \left( \begin{array}{cccc}
   1 & 0  & 0 & 0 \\ 0 & 1 & 0 & 0 \\ 0 & 0 & -1 & 0 \\ 0 & 0 & 0 & 3
   \end{array} \right) ,
\label{c38}
\end{equation}
to conforming with the super-tracelessness of the $SU(2/2)$ generators.
The superconnection for the $SU(2/2)$ case can be written as
%eq.39
\begin{equation}
\cj\
=it_i J_i \ \ (i=1,2,\cdots ,15)
= \left( \matrix{A_L + {\frac{1}{\sqrt{2}}} B&i\Phi \cr
i{\Phi}^{\dag} &A_R +
{\frac{1}{\sqrt{2}}} B\cr} \right)
\label{c39}
\end{equation}
with one-forms in the even part and zero-forms in the odd part, given as
%eq.40
\begin{equation}
A_L =i\tau_a A_{La} ,\ \  A_R =i\tau_a A_{Ra} , \ \ B=i I Y ,\ \
\Phi =I\phi_0 + i\tau_a \phi_a ,
\label{c40}
\end{equation}
where $\tau_a (a=1,2,3)$ are Pauli matrices, and $I$ is $2 \times 2$
identity matrix. $A_{La}, A_{Ra}, Y$ are real, whereas $\phi_0 , \phi_a$
are complex, the fields being assigned to the components of $J$'s
according to
\begin{eqnarray*}
A_{La}& =& J_{a} \   (a=1,2,3), \ \ A_{R_{1}}=J_{13},
 \ \ A_{R_{2}} = J_{14},\\
 A_{R_{3}}& =& - \frac{1}{\sqrt{3}}(J_{8} + \sqrt{2} J_{15}), \ \
 Y = - \frac{1}{\sqrt{3}}(\sqrt{2} J_{8} - J_{15}), \\
\phi_{0}& =& \frac{1}{2}[(J_{4} - i J_{5}) + ( J_{11} - i J_{12})], \ \
\phi_{1} = -\frac{i}{2}[(J_{6} - iJ_{7}) + (J_{9} - iJ_{10})], \\
\phi_{2}& =& -\frac{1}{2}[(J_{6}- iJ_{7}) - (J_{9} - i J_{10})], \ \
\phi_{3} = -\frac{i}{2}[(J_{4} - i J_{5}) - (J_{11} - i J_{12})] .
\end{eqnarray*}
$A_L$ and $A_R$ are thus the $SU(2)$ gauge fields, $B$ is the $U(1)$ gauge
field, and $\Phi$ is the complex scalar field (with its four components).

We now introduce the ghost and anti-ghost fields,
%eq.41
\begin{equation}
\bc\ =\left(\matrix{c_L&0\cr 0&c_R\cr}\right)
+\left(\matrix{{\frac{1}{\sqrt{2}}}c_I&0\cr 0&
{\frac{1}{\sqrt{2}}}c_I\cr}\right), \;
\bcb =\left(\matrix{\bar{c_L}&0\cr 0&\bar{c_R}\cr}\right)
+\left(\matrix{{\frac{1}{\sqrt{2}}}\bar{c_I}&0\cr 0&
{\frac{1}{\sqrt{2}}}\bar{c_I}\cr}\right),
\label{c41}
\end{equation}
where $c_L =i\tau_a c_{La}, c_R =i\tau_a c_{Ra} (a=1,2,3), c_I =iIc_{Ir},$
with real $c_{La}, c_{Ra}$, and $c_{Ir}$ and similarly for $\bcb\ $.
$\{ c_L , {\bar{c_L}} \}$ and $\{ c_R , {\bar{c_R}} \}$ are the ghost and
antighost fields for the $SU(2)$ gauge fields $A_L$ and $A_R$, respectively,
and $\{ c_I , {\bar{c_I}} \}$ are those of the $U(1)$ gauge field $B$.

The BRST/anti-BRST transformation rules are obtained from
(\ref{c15})-(\ref{c18}). Choosing
%eq.42
\begin{equation}
 \eta =\left( \begin{array}{cc}
0 &\xi \\ \xi^\dagger &0 \end{array} \right), \; {\rm where}
\; \; \xi =k\left( \begin{array}{cc}
1 & 0\\ 0&1 \end{array}\right), \; k: \ {\rm real},
\label{c42}
\end{equation}
we get
%eq.45
\begin{eqnarray}
s A_L & = & -d c_L - A_L c_L -c_L A_L , \nonumber\\
\bar{s} A_L & = & -d \bar{c_L} - A_L \bar{c_L} - \bar{c_L} A_L , \nonumber \\
s A_R & = & -d c_R - A_R c_R -c_R A_R , \nonumber\\
\bar{s} A_R & = & -d \bar{c_R} - A_R \bar{c_R} - \bar{c_R} A_R , \nonumber \\
s B & = & -d c_I , \ \  \bar{s} B = -d \bar{c_I} , \nonumber \\
s \Phi & = & (\Phi +\xi ) c_R -c_L (\Phi +\xi ) , \nonumber\\
\bar{s} \Phi &=& (\Phi +\xi )\bar{c_R} - \bar{c_L} (\Phi +\xi ) , \nonumber \\
s \Phi^{\dag} & = & (\Phi^{\dag} +\xi^{\dag} ) c_L -c_R (\Phi^{\dag}
 +\xi^{\dag} ) , \nonumber\\
\bar{s} \Phi^{\dag} &=& (\Phi^{\dag} +\xi^{\dag} )\bar{c_L}
- \bar{c_R} (\Phi^{\dag} +\xi^{\dag} ) , \nonumber \\
s c_L & = & -c_Lc_L ,  \ \
\bar{s} \bar{c_L} = -\bar{c_L} \bar{c_L}
\label{c43} , \\
s c_R & = & -c_Rc_R ,  \ \
\bar{s} \bar{c_R} = -\bar{c_R} \bar{c_R}
\nonumber ,\\
s c_I & = & \bar{s} \bar{c_I} = 0
\nonumber , \\
s \bar{c_L} & = & b_L , \ \
\bar{s} c_L = -b_L -c_L\bar{c_L} - \bar{c_L} c_L , \nonumber \\
s \bar{c_R} & = & b_R , \ \
\bar{s} c_R = -b_R -c_R\bar{c_R} - \bar{c_R} c_R , \nonumber \\
s b_L & = & 0  , \ \
\bar{s} b_L = - \bar{c_L} b_L + b_L\bar{c_L} , \nonumber\\
s b_R & = & 0  , \ \
\bar{s} b_R = - \bar{c_R} b_R + b_R\bar{c_R}
\nonumber , \\
s \bar{c_I} & = & - \bar{s} c_I = b_I , \ \
s b_I = {\bar{s}} b_I = 0.
\nonumber
\end{eqnarray}
We have introduced the auxiliary fields
$\cee\ =\left(\matrix{b_L&0\cr 0&b_R\cr}\right) +
\left(\matrix{{\frac{1}{\sqrt{2}}}b_I&0\cr 0&
{\frac{1}{\sqrt{2}}}b_I\cr}\right)$
with $b_L =i\tau_a b_{La}$, $b_R =i\tau_a b_{Ra} \ \  (a=1,2,3),$
and $b_I = i I b_{Ir}$, where $b_{La} , b_{Ra}$, and $b_{Ir}$ are real.
For the supercurvature, we obtain
%eq.44
\begin{equation}
\cf\ =
=\left( \matrix{F_L +{\frac{1}{\sqrt{2}}}F_B -(\Phi {\Phi}^{\dag}
+\xi {\Phi}^{\dag} +\Phi {\xi}^{\dag} )& \ \ -i
(D\Phi + A_L \xi -\xi A_R )
\cr -i
(D {\Phi}^{\dag} - {\xi}^{\dag} A_L + A_R {\xi}^{\dag} )
& \ \ F_R +{\frac{1}{\sqrt{2}}}F_B
-({\Phi}^{\dag} \Phi + {\xi}^{\dag} \Phi + {\Phi}^{\dag} \xi )\cr}
\right) ,
\label{c44}
\end{equation}
where
$F_L=dA_L+A_LA_L , \ \ F_R=dA_R+A_RA_R , \ \ F_B=dB , \ \
D \Phi =d\Phi +A_L \Phi
-\Phi A_R , \ \ D {\Phi}^{\dag} =d{\Phi}^{\dag}
-{\Phi}^{\dag} A_L + A_R {\Phi}^{\dag}$.

The classical Lagrangian, the first term in (\ref{c23}), is given by
\cite{thl},
%eq.45
\begin{eqnarray}
{\cal L}_C &=&tr[\frac{1}{4}F_{+\mu\nu} F_{+}^{\mu\nu} +
                \frac{1}{4}F_{-\mu\nu} F_{-}^{\mu\nu} +
                \frac{1}{8}F_{B\mu\nu} F_{B}^{\mu\nu}   \nonumber\\
&-&\frac{1}{2}(D {\Phi}^{\dag} + 2kA_-)_{\mu}(D \Phi - 2kA_-)^{\mu}
-\frac{1}{2}({\Phi}^{\dag}\Phi  + k(\Phi +{\Phi}^{\dag}))^2 ],
\label{c45}
\end{eqnarray}
where $A_{\pm}$ are respectively the vector and axial vector gauge fields,
as defined by
$A_{\pm}=\frac{1}{2}(\pm A_L + A_R)=i{\tau}_aA_{\pm a}$, and
$F_{+}^{\mu\nu}={\partial}^{\mu}A_{+}^{\nu}
             - {\partial}^{\nu}A_{+}^{\mu}
             + [A_{+}^{\mu},A_{+}^{\nu}]
             + [A_{-}^{\mu},A_{-}^{\nu}],\ \
F_{-}^{\mu\nu}={\partial}^{\mu}A_{-}^{\nu}
             - {\partial}^{\nu}A_{-}^{\mu}
             + [A_{+}^{\mu},A_{-}^{\nu}]
             + [A_{-}^{\mu},A_{+}^{\nu}].$
The above expression tells us that the three axial vector gauge fields
$A_{-a}$ have acquired the mass $2k$, whereas the three vector gauge fields
$A_{+a}$ and the $U(1)$ gauge field $Y$ remain massless.

For the quantum Lagrangian
${\cal L}_Q$,
we again write, as in (\ref{c34}),
\[{\cal L}_Q = {\cal L}_C + {\cal L}_1 + {\cal L}_2 .\]
The ghost part ${\cal L}_1$ is given by
%eq.46
\begin{eqnarray}
{\cal L}_1 &=& {\frac{1}{2}}
tr[({\partial}_{\mu} \bar{c_L} D^{\mu} c_L
 +{\partial}_{\mu} \bar{c_R} D^{\mu} c_R
 +{\partial}_{\mu} \bar{c_I} {\partial}^{\mu} c_I )
 -2k^2 (\bar{c_L} -\bar{c_R} ) (c_L -c_R )\label{c46}\\
 &+&k
(\{ (\bar{c_L} -\bar{c_R} ) c_R - c_L (\bar{c_L} -\bar{c_R} )\}{\Phi}^{\dag}
+ \{ c_R (\bar{c_L} -\bar{c_R} ) - (\bar{c_L} -\bar{c_R} ) c_L \} \Phi ) ] ,
\nonumber
\end{eqnarray}
where $D^{\mu} c_L = {\partial}^{\mu} c_L + [ A_L^{\mu} , c_L ] ,\ \
D^{\mu} c_R = {\partial}^{\mu} c_R + [ A_R^{\mu} , c_R ] .$ \\
The gauge fixing part ${\cal L}_2$ is given by
%eq.47
\begin{equation}
{\cal L}_2 =  \alpha \ \  \{ ((b_{-a})^2 + (b_{+a})^2 +
{\frac{1}{2}} (b_{Ir})^2 )
+\frac{2}{\alpha} (b_{-a} ({\partial}_{\mu} A_{-a}^{\mu} - 2k{\varphi}_a )
+b_{+a} ( {\partial}_{\mu} A_{+a}^{\mu} ) +
{\frac{1}{2}} b_{Ir} ( {\partial}_{\mu} Y^{\mu} ) ) \} ,
\label{c47}
\end{equation}
where $b_{\pm} = \frac{1}{2} (\pm b_L + b_R) = i {\tau}_a b_{\pm a} ,\ \
\varphi = \frac{1}{2} (\Phi -{\Phi}^{\dag})={\tau}_a{\varphi}_a .$
Integrating out the auxiliary fields $b_{\pm}$, $\; {\cal L}_2$ becomes
%eq.48
\begin{equation}
{\cal L}_2 = - {\frac{1}{\alpha}} \ \  \{ ({\partial}_{\mu} A_-^{\mu} -2k
\varphi )^2 + ( {\partial}_{\mu} A_+^{\mu} )^2
+ {\frac{1}{2}} ( {\partial}_{\mu} Y^{\mu} )^2 \} .
\label{c48}
\end{equation}
This expression again displays the quantum Lagrangian in the 't Hooft gauge:
%eq.49
\begin{eqnarray}
{\partial}_{\mu} A_{-a}^{\mu} - M_{A_-} {\varphi}_a &=& 0 , \nonumber\\
{\partial}_{\mu} A_{+a}^{\mu} &=& 0 , \label{c49}\\
{\partial}_{\mu} Y^{\mu} &=& 0 , \nonumber
\end{eqnarray}
where $M_{A_-}=2k$ is used.
If we write $\Phi$ as $(\sigma + i \vec{\pi}\cdot\vec{\tau} )
+i(\eta + \vec{\rho}\cdot\vec{\tau} )$ with real $\sigma , \vec{\pi} ,
\eta ,\vec{\rho}$ fields, then $\varphi$ in ${\cal L}_2$
can be identified with $\vec{\pi}$.
This is consistent with the fact that the $\vec{\pi}$ fields are gauged
away and mutate into the longitudinal components of the axial vector fields
$A_{-}$ in the unitary gauge. This is also related to the fact that the
$SU(2/2)$ case corresponds to the gauged $SU(2) \times SU(2)$
$\sigma$-model \cite{thl}.\\

\noindent
{\bf \large V. Conclusion}\\

In the matrix derivative approach, derived from noncommutative geometrical
gauge theory and adjoined to internal supersymmetry, in its superconnection
version, the vector gauge fields and the scalar fields are combined together,
constituting the superconnection. The two sets of fields are thus related
as a supermultiplet from the very beginning. This provides for an elegant
geometrical realization of the Higgs mechanism. The entire Lagrangian is
geometrical, even including the negative mass term for the scalar field,
needed to trigger the spontaneous symmetry breakdown for the (g-even) gauge
subgroup. That symmetry-breaking quadratic term for the scalar field is
provided by the matrix derivative, beyond the unification achieved by the
supergroup by itself. Summarizing, the unification is complete, within
the limitations set by the broken symmetry actual content. We return to
these limitations in our last paragraph.

Another advantage of the formalism touches upon the quantum action, namely
in the gauge in which it appears, as a result of the construction. This turns
out to be the `t Hooft gauge, most convenient for a spontaneously broken
symmetry with Higgs field and suitable for renormalization \cite{tht,fls}.
We obtained this action just by adapting the
Baulieu/Thierry-Mieg method \cite{btm},
which would yield the Landau gauge for the unbroken Yang-Mills theory,
 to the matrix derivative approach.

For the calculation of the ${\cf\ }^{\star}\cdot \cf\ $ term in the
classical and some of the other parts of the quantum Lagrangian
we have used the definition of (\ref{c31}) for the dual form. This
definition gives the kinetic terms of both the vector and scalar fields
automatically in their canonical form, also providing the relation
$M_{\chi} =2M_W$.\footnote{This is a classical relationship. Including
quantum correction should modify this result. For a related result, see
Ref. \cite{hln}, in an application of the superconnection approach without
introducing the notion of a matrix derivative.}
This ratio is also due to the fact that we have only one overall supergauge
coupling constant $g$ for the superconnection $\cj\ $ in section 3, due to
universality. Without the assumption of universality for the supergroup we
would have independent couplings for fields corresponding to forms of
different degrees - in our case the even and odd parts of the superconnection,
i.e. two independent couplings. One might then obtain a different mass ratio
for the Higgs and gauge bosons \cite{coqsch}.

Lastly, we note that only the even part of the supergroup is gauged in the
sense of Relativistic Quantum Field Theory - even though the entire supergroup
is used as a structure group for the theory and provides the geometrical
framework for the quantization procedure, including the `t Hooft gauge. As
a result, there is no guarantee of non-renormalization of the theory's
couplings beyond those of the g-even gauge subgroup.\\

\noindent
{\em Acknowledgements} \\
\indent
CYL would like to thank Hoil Kim for helpful discussions on the mathematics
of noncommutative geometry.
This work was supported in part by NON DIRECTED RESEARCH FUND, Korea
Research Foundation, in part by the KOSEF through the SRC program of
SNU-CTP, and in part by the BSRI program of the Ministry of Education.\\


\begin{thebibliography}{99}

\bibitem{yn} Y. Ne'eman, Phys. Lett. B81 (1979) 190. For a different
  approach, see also D. B. Fairlie, Phys. Lett. B82 (1979) 97.
\bibitem{nt} Y. Ne'eman and J. Thierry-Mieg, Proc. Nat. Acad. Sci.
USA 77 (1980) 720.
\bibitem{ntm} Y. Ne'eman and J. Thierry-Mieg, in Proceedings of the
  Salamanca(1979) International Conference on Differential Geometric
  Methods in Mathematical Physics, ed. A. Perez-Rendon,
  Lecture Notes in Mathematics No. 836 (Springer-Verlag, Berlin,
  1980).
\bibitem{ntm2} Y. Ne'eman and J. Thierry-Mieg, Phys. Lett. B108
     (1982) 399.
\bibitem{ns} Y. Ne'eman and S. Sternberg, Proc. Natl. Acad. Sci.
              USA 87 (1990) 7875.
\bibitem{ln} C.Y. Lee and Y. Ne'eman, Phys. Lett. B264 (1991) 389;
          {\sl ibid.} B269 (1991) 477.
\bibitem{conn} A. Connes, in "The interface of mathematics and
    particle physics", ed. D. Quillen, G. Segal and S. Tsou
    (Oxford Univ. Press, Oxford, 1990).
\bibitem{conlot} A. Connes and J. Lott,
    Nucl. Phys. B (Proc. Suppl.) 18B (1990) 29.
\bibitem{chamff1} A. H. Chamseddine, G. Felder and J. Fr\"{o}hlich,
   Nucl. Phys. B395 (1993) 672.
\bibitem{chamff2} A. H. Chamseddine, G. Felder and J. Fr\"{o}hlich,
  Comm. Math. Phys. 155 (1993) 205.
\bibitem{cham} A. H. Chamseddine, ``Connection between space-time supersymmetry
      and non-commutative geometry", hepth-9404138.
\bibitem{couqet} R. Coquereaux, G. Esposito-Far\`{e}se and G. Vaillant,
   Nucl. Phys. B353 (1991) 689.
\bibitem{sch} R. H\"{a}u{\ss}ling, N.A. Papadopoulos and F. Scheck,
        Phys. Lett. B260 (1991) 125.
\bibitem{qui} D. Quillen, Topology 24 (1985) 89.
\bibitem{pps} N.A. Papadopoulos, J. Plass, and F. Scheck, Preprint MZ-TH/93-26
              (1993).
\bibitem{hln} D.S. Hwang, C.Y. Lee and Y. Ne'eman,
  ``BRST quantization of SU(2/1) electro-weak theory in the superconnection
  approach - and the Higgs meson mass",
  to be published in Int. J. Mod. Phys. A.
\bibitem{tm} J. Thierry-Mieg, J. Math. Phys. 21 (1980) 2834.
         For a doubled fiber bundle structure, see
             M. Quiros, F. J. De Urries, J. Hoyos, M. L. Mazou, and
             E. Rodriguez, {\sl ibid.} 22 (1981) 1767 and the next reference.
\bibitem{tmn} J. Thierry-Mieg and Y. Ne'eman, Proc. Natl. Acad. Sci.
         USA 79 (1982) 7068.
\bibitem{btm} L. Baulieu and J. Thierry-Mieg, Nucl. Phys. B197
      (1982) 477.
\bibitem{tht} G. 't Hooft, Nucl. Phys. B33 (1971) 173;
             {\sl ibid.} B35 (1971) 167.
\bibitem{fls} K. Fujikawa, B. W. Lee and A. I. Sanda,
   Phys. Rev. D6 (1972) 2923.
\bibitem{egh} T. Eguchi, P. Gilkey and A. Hanson,
   Phys. Rep. 66 (1980) 213.
\bibitem{thl} D. S. Hwang and T. Lee, Int. J. Mod. Phys. A9 (1994) 5531.
\bibitem{coqsch} R. Coquereaux, G. Esposito-Far\`{e}se and F. Scheck,
   Int. J. Mod. Phys. A7 (1992) 6555.
\end{thebibliography}
\end{document}